\documentclass[12pt]{article}
\usepackage{setspace}
\usepackage{amsfonts}
\usepackage{amsmath}
\usepackage{amssymb}
\usepackage{amsthm}
\usepackage{mathrsfs}
\usepackage{graphics}
\usepackage{lscape}
\usepackage{changebar}
\usepackage{graphicx}
\usepackage{epsfig}
\usepackage{booktabs}
\usepackage{enumerate}
\usepackage{url}
\usepackage{caption}
\usepackage{natbib}
 \renewcommand\refname{\normalsize\bf Literature Cited }

\renewcommand{\vec}[1]{\mathbf{#1}}
\newcommand{\taub}{\mbox{\boldmath$\tau$}}

\newcommand{\deriv}[2]{\frac{\partial #1}{\partial #2}}

\newcommand{\sigb}{\mbox{\boldmath$\sigma$}}

\newcommand{\gdotb}{\mbox{\boldmath$\dot\gamma$}}

\makeatother

\usepackage{color}
\definecolor{cyan}{cmyk}{1,0,0,0}
\definecolor{black}{cmyk}{0,0,0,1}
\definecolor{magenta}{cmyk}{0,1,0,0}
\definecolor{yellow}{cmyk}{0,0,1,0}
\definecolor{blue}{cmyk}{1,1,0,0}
\definecolor{red}{cmyk}{0,1,1,0}
\definecolor{green}{cmyk}{1,0,1,0}

\begin{document}

\title{\bf \large Symmetry relations in viscoplastic drag laws}

\author{ \normalsize K. Kamrin$^*$ and J. Goddard$^{
**}$
\\\small
$^*$Department of Mechanical Engineering\\\small
Massachusetts Institute of Technology\\\small
$^{**}$Department of Mechanical and Aerospace Engineering\\ \small University of California, San Diego}
%9500 Gilman Drive,  La Jolla, CA 92093-0411
%}
%\volume{0}
\date{} 
\maketitle

\section*{\normalsize Abstract} The following note shows that the symmetry of various   resistance formulae, often  based on Lorentz reciprocity for linearly viscous fluids,  applies to a  wide class of non-linear viscoplastic fluids. This follows from Edelen's non-linear  generalization of the Onsager relation for  the  special case of \emph{strongly dissipative} rheology, where constitutive equations   are derivable from his dissipation potential. 
For flow domains with strong dissipation in the interior and on a portion of the boundary this implies strong dissipation on the remaining portion of the boundary, with strongly dissipative traction-velocity response given by a dissipation potential. This  leads to a non-linear  generalization of Stokes resistance formulae for a wide class of viscoplastic fluid problems.  We consider the application to non-linear Darcy flow and to the effective slip for viscoplastic flow over textured surfaces.

\section*{\normalsize Introduction}
Symmetry occupies an important position in the classical linear theories of elasticity, viscosity  and viscoelasticity,  where it is synonymous with   self-adjointness
and Lorentz reciprocity. 

In the case of (hyper)elastic materials, the symmetry of the linear-elastic modulus is a consequence of the existence of
 a strain-energy function, whereas in the case of linearly viscous  fluids, the symmetry of the viscosity tensor represents Rayleigh-Onsager
 symmetry. As shown by Day \citep{Day71}, the same symmetry applies 
 to the  linear-viscoelastic memory function for materials that exhibit time-reversibility on certain closed strain paths\footnote{It is not too difficult to show that for general linear-response theory,  this implies a Hermitian matrix $\hat{\vec{L}}(\omega)$ in  the  Fourier description  $\hat{ \vec{f} }(\omega) = \hat{\vec{L}}(\omega)\hat{\vec{v}}(\omega)$ in the frequency domain
of the linear relation connecting generalized force $\vec{f}(t)$ and velocity $\vec{v}(t)$ in the time domain.}.

For the special case of Newtonian fluids there have been numerous applications of 
Lorentz reciprocity to various problems of Stokes (inertialess) flow \citep{Happel65,Hinch1972,Brunet2004} to obtain various symmetry restrictions on 
the associated drag laws.  In a similar spirit,  this same principle has 
 been applied to the symmetries of  apparent slip in the  far-field 
above arbitrarily textured surfaces with arbitrary local Navier slip distribution \citep{Kamrin11}.  This problem area has received attention in recent day due to applications in microfluidics; see references in \citep{kamrin2010} for a large body of related work. 

While the above symmetry is bound up with variational principles based on 
the associated quadratic forms,  there exist more general nonlinear variants.   
Thus, in  the case of non-linear hyperelasticity, there exist well-known elastostatic variational principles,  with elastic stress given by the gradient of a strain-energy function,  or by an  associated pseudo-linear form involving a symmetric (tangent)
modulus based on the Hessian of a  complementary energy. 

Less well known are the analogous forms for strictly dissipative  nonlinear systems
given by the general theory of Edelen \citep{Edelen72, Edelen73, Goddard14} as  
a generalization of Onsager symmetry. 
In particular,  Edelen proves that,  modulo a gyroscopic or ``powerless" 
force, the dissipative force $\vec{f}$ is given as the gradient $\partial_{\vec{v}}\psi(\vec{v})$ of a dissipation potential $\psi(\vec{v})$ depending on a generalized
velocity $\vec{v}$ or, again, by a pseudo-linear form based on the Hessian of 
a complimentary potential.  Whenever the gyroscopic force is identically zero, we 
call the system \emph{strongly dissipative} or, by analogy to the 
elastic case, \emph{hyperdissipative}. For later reference, we note the 
dual form $\vec{v}=\partial_{\vec{f}}\varphi(\vec{f})$ where $\varphi$ is the (Legendre-Fenchel) convex conjugate of  $\psi$. 

The main goal of the current note is to identify and exploit connections between strongly dissipative \emph{local} properties of a system (i.e. constitutive relations and/or surface interactions) and strongly dissipative global properties, which often take the form of homogenized macroscale relations.  In particular, we will demonstrate how this result restricts (i) Darcy-like laws for porous flow, and (ii) effective slip relations over textured surfaces, when the fluid rheology and surface interactions are non-linear while maintaining a strongly dissipative form.  Throughout, we emphasize the ubiquity of strong dissipation among commonly used viscoplastic fluid models and, hence,  the generality of the  results to be presented.

As discussed elsewhere \citep{Goddard14}, Edelen's work has interesting 
implications for  rate-independent rigid plasticity, where plastic potentials are often based on  rather special physical arguments.  Moreover, Edelen's work provides a rigorous mathematical extension to the phenomenological viscoplastic  potentials  introduced by others \citep{Rice70}, which leads to interesting analogies  between viscoplasticity  and non-linear elasticity.  In both domains, 
variational principles  govern quasi-static  stress equilibrium, and the  associated symmetry carries over to various symmetries in global 
force laws. Some existing applications of highly special viscoplastic potentials involve extremum principles  for non-Newtonian flow \citep{Johnson61} and bounds  for nonlinear homogenization \citep{Dormieux06}.  

The purpose of the present work is to introduce variational forms into strongly dissipative flow problems and to explore the consequences for some special cases involving viscoplastic flow within porous 
media and over textured surfaces.  In each case, a set of symmetry conditions and differential constraints arise as necessary conditions on the homogenized resistance formulae.  %\jcom{Ken, I am not sure of the distinction between symmetry and "compatability" in the present context. Since "compatability" has several interpretations, is there some risk of confusion in using it here? Also, do we really have to point to specific equations at this point? A set of symmetry and compatibility requirements arise as necessary conditions on the homogenized permeability (Eq \ref{porous_compat}) and surface mobility relation (Eq \ref{mob_compat}).}

\section*{\normalsize Strongly dissipative rheology and mobile boundaries}

In what follows  $\sigb$ represents Cauchy stress, $\vec{v}$ material velocity, and $\vec{D}=\text{sym}(\nabla \vec{v})$ associated strain-rate tensor, all fields depending on spatial position and time, $\vec{x}$ and $t$.   We employ the prime ${}^\prime$ to denote the deviator, with  \begin{equation}\vec{S}\equiv\sigb^\prime\equiv\sigb-\frac{1}{3}(\text{tr}\: \sigb)\vec{1}, \:\:\mbox{and}\:\:\vec{E}=\vec{D}^\prime={\bf D}-\frac{1}{3}(\text{tr}\: \vec{D})\vec{1},\end{equation}  where $\vec{1}$ denotes the three-dimensional identity tensor.

We begin by defining a strongly dissipative incompressible viscoplastic fluid as one which obeys
\begin{equation}\label{const}\begin{split}&
\vec{S}=\vec{S}(\vec{E})=\partial_{\vec{E}}\psi(\vec{E})^\prime, \:\:\mbox{with}\:\: \mathfrak{D}=\vec{E}\!:\:\vec{S}(\vec{E})\ge0, \\&\mbox{and}\:\:
 \mathfrak{D} >0 \:\:\mbox{for}\:\: |\vec{E}|\ne 0 \:\:\& \:\:
\mathfrak{D}=0\:\: \mbox{for}\:\: |\vec{E}|= 0.
\end{split} \end{equation}
 $\mathfrak{D}$ denotes dissipation (per volume), and  the inequality represents a convexity condition on $\psi$.  Here as below, the
colon $:$ denotes tensorial contraction, with e.g.\ $\vec{A}\!:\!\vec{B} = $\:tr $ \vec{A}\vec{B}^T$ representing the (Euclidian) scalar product of real tensors. Also, we write   
$\partial_X=  (\partial\:/\partial X)^T$ and the tensorial derivative carries the usual definition, i.e. $[\partial_{\vec{E}}\psi(\vec{E})]_{ij}=\partial_{{E}_{ij}}\psi(\vec{E})$.

The function  $\psi$ represents a \emph{dissipation potential} according to the definition of Edelen. Owing to its convexity,  the roles of the dependent and independent variables can be swapped, since  there exists a dual potential or Legendre convex conjugate $\varphi(\vec{S})=\vec{S}\!:\:\vec{E}(\vec{S})-\psi(\vec{E}(\vec{S}))$,  such that the  inverse of (\ref{const}) is given by 
\begin{equation}\label{dual}
\vec{E} =\vec{E}(\vec{S})=\partial_{\vec{S}}\varphi(\vec{S})^\prime, \:\:\mbox{with}\:\:\vec{S}\!:\:\vec{E}(\vec{S})\ge 0
\end{equation}
We note that this duality  is singular in the case of certain non-smooth rheologies, 
such as rate-independent plasticity, where $\psi(\vec{E})$ is a homogeneous function of degree one \citep{Goddard14}. 
%We note that non-invertible or non-smooth rheologies may possess corresponding ill-posedness of $\phi$ or $\psi$. 

It turns out that many common non-Newtonian models have the strongly dissipative form. As examples, Table \ref{Table:common} lists some standard models of viscoplastic fluids, 
 %eall of which are special cases of the generalized Newtonian fluid:
%\begin{equation}\label{common} \vec{S}=2\eta(|\vec{E}|)\vec{E}, \:\:\mbox{with}\:\:\psi= \int_0^{|%\vec{E}|}\eta(s)s\ ds, \end{equation}
with  $|\vec{E}|=(\vec{E}\!:\!\vec{E})^{1/2}$.  The Bingham and Herschel-Bulkley models represent a class of ``yield-stress fluids", which have indeterminate stress at 
the rest state $\vec{E}=\vec{0}$, unless the unit director $\vec{E}/|\vec{E}|$ is specified. The long-standing problem of determining the spatial
location of yield surfaces is the subject of ongoing research cited in a  recent review article \citep{Denn11}.

\begin{table}[t]
\begin{center} 
\begin{tabular}{lll}
\toprule
\multicolumn{1}{l}{Fluid Model} & 
\multicolumn{1}{l}{Deviatoric stress $\vec{S}$}&\multicolumn{1}{l}{Dissipation potential $\psi$}\\
\midrule
 Newtonian & $ 2\eta\vec{E}$ &  $ \eta|\vec{E}|^2$\\
Power-law & $ 2K |\vec{E}|^{n-1} \vec{E}$& $2K \left|\vec{E}\right|^{(n+1)}\!\!/(n+1)$% & $\frac{ |(\sigb'/2K)|^{1+1/n}}{1+1/n}$
\\
Bingham plastic&  $\mu\vec{E}/|\vec{E}|+2\eta\vec{E}$& $\mu |\vec{E}|+\eta|\vec{E}|^2$
\\
Herschel-Bulkley& $ \mu\vec{E}/|\vec{E}|+2K|\vec{E}|^{n-1}\vec{E}$& $\mu|\vec{E}|+2K|\vec{E}|^{1+n}/(1+n)$
\\
\bottomrule
\end{tabular}
\caption{Common incompressible  fluid models exhibiting a strongly dissipative form}\label{Table:common}
\end{center}
\end{table}

%Eq.~\ref{common} subsumes many commonly employed rheological models and is 

The models displayed in the table are special cases of a potential $\psi(\vec{E})=\Psi(I_2, I_3)$,  where  $I_2= |\vec{E}|^2/2$ and $I_3=$ det $\vec{E}$ are  the non-zero isotropic invariants of $\vec{E}$.  This potential  yields the incompressible isotropic (Reiner-Rivlin) model:
\begin{equation}\label{RR}
\vec{S}=(\partial_{I_2}\Psi)\, \vec{E}+(\partial_{I_3}\Psi)\, (\vec{E}^2-\frac{1}{3}|\vec{E}|^2\vec{1})=\boldsymbol{\eta}(\vec{E})\!:\!\vec{E}, 
\end{equation}
where $\boldsymbol{\eta}(\vec{E})=[\eta_{ijkl}(\vec{E})]$ is a non-negative fourth-rank viscosity tensor representing a {\it pseudo-linear} form \citep{Goddard14} of a 
type that appears frequently in the following.

We recall that the dissipation potentials  in Table \ref{Table:common}
follow from the special forms considered by Hunter  \citep{Hunter76}, who may have been  unaware of the general theory of Edelen, and we note  that certain variational principles have been formulated for the special case of the ``generalized Newtonian fluid" 
\citep{Johnson61} $\vec{S}= \eta(|\vec{E}|)\vec{E}$ involving a scalar viscosity depending on a single scalar invariant. 

 All the above isotropic fluid  models are special cases of a more 
general \emph{anisotropic}  fluid,  with $\sigb=\partial_{\vec{D}}\psi(\vec{D}, \boldsymbol{\mathcal{S}})=\boldsymbol{\eta}(\vec{D}, \boldsymbol{\mathcal{S}})\!\!:\!\!\vec{D}$, where Edelen's dissipation
potential $\psi$ depends on the joint isotropic invariants of $\vec{D}$ and a
set of ``structure tensors" $\boldsymbol{\mathcal{S}}$. Moreover, such tensors may depend on the history of flow, with evolution  described by a set of objective Lagrangian ODEs depending on $\vec{D}$. For example,
in the case of a single ``fabric" tensor $\vec{A}$, the evolution equation 
takes the form 
\begin{equation}\label{fabric} \stackrel{\circ}{\vec{A}}= \boldsymbol{\mathfrak{a}}(\vec{A}, \vec{D}).\end{equation}
where superposed ``$\circ$" denotes the Jaumann or corotational derivative. The joint 
isotropic invariants of $\vec{A}, \vec{D}$ are well known, and are represented
by a finite set of traces of the form tr$(\vec{A}^n\vec{D}^m)$.  Models of this type also
allow for change of density and include Reiner's ``dilatant" isotropic fluid \citep{Reiner45} as well
various anisotropic variants with application to dilatant granular media.
That said, we focus attention in the present work on incompressible fluids.

It is worth recalling several previous works on  non-linear flow in porous  media, including less general models of non-Newtonian flow in Bear \citep[p.\ 128]{Bear88} and Dormieux et al. \citep{Dormieux06} as well as  turbulent flow of Newtonian fluids, where one can also define a dissipation potential.  See e.g.  \cite{Joseph82} or  \citep[p.\ 177]{Bear88}  and \cite{Dormieux06}.

\subsection*{\it \normalsize Boundary conditions}

\begin{figure}[h]
\begin{center}
\includegraphics[width=6cm]{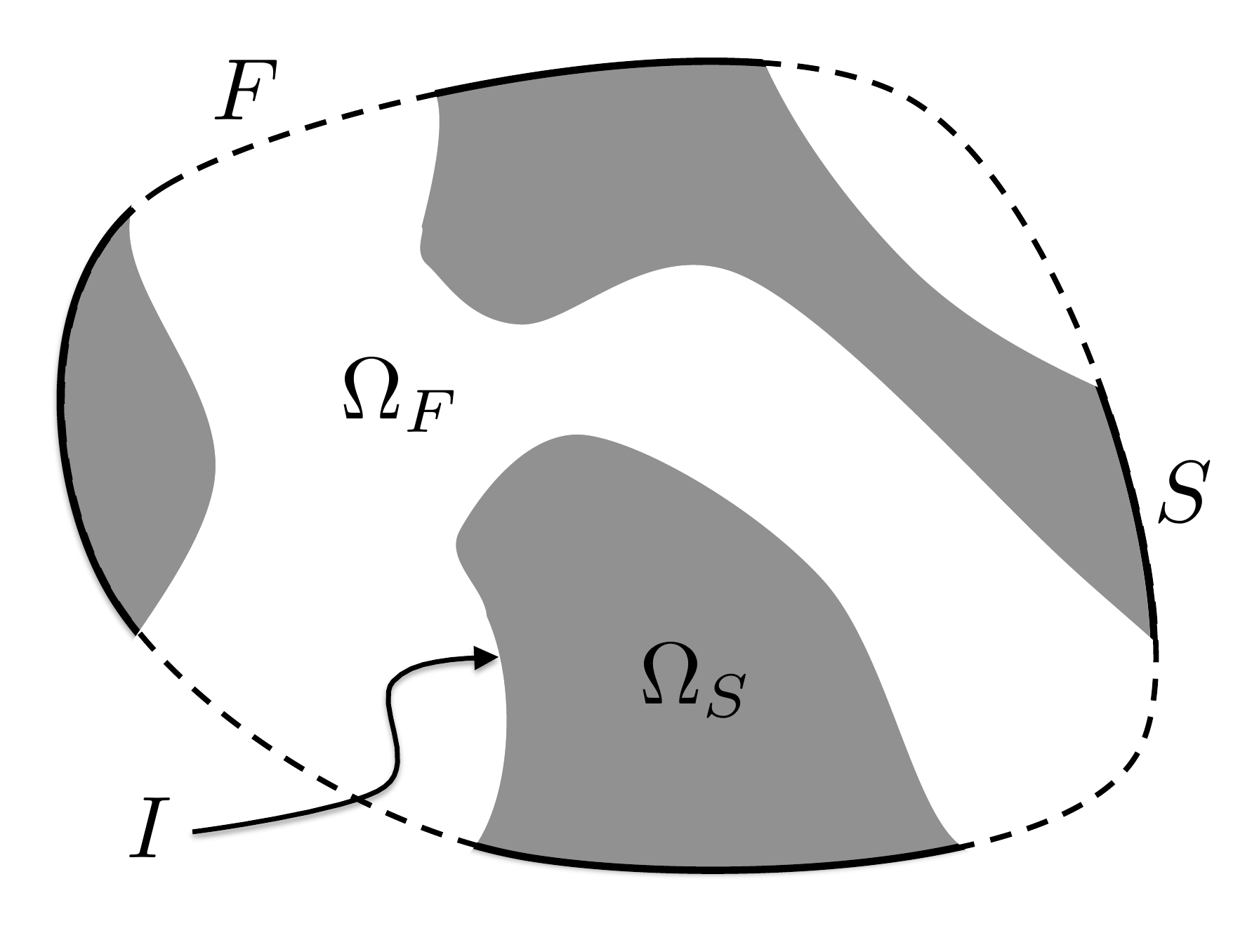}
\caption[caption]{ 2D schematic of  flow in a porous medium $\Omega=\Omega_S\cup\Omega_F$ or in a (simply connected) fluid region $\Omega_F$ above a textured surface $I$ (See Fig. \ref{Fig:flow}).}
\label{Fig:general1}
\end{center}
\end{figure}
We turn now from bulk constitutive relations to a consideration of boundary conditions.  Thus, we imagine a 
region of space $\Omega=\Omega_F \cup \Omega_S$ with $I \cap F=\varnothing$, consisting of disjoint
surfaces $I$ and $F$, where $I$ represents the interface between the fluid and  solid region and  $F$ is the bounding surface lying in the fluid.  Then, the boundary of the fluid domain is $\partial\Omega_F =I \cup F$.

For example, in the porous solid illustrated in  Fig.~\ref{Fig:general1}, $F$  represents that part of $\partial\Omega$ lying in the fluid, with  $I$ representing the interface between $\Omega_S$ and $\Omega_F$  on which there is partial slip with no permeation.  It may be possible to extend our analysis so as to permit permeation through $I$,   but we shall  specialise shortly to the case of an impermeable surface $I$.

% Within a continuum-level description of the porous medium  we may regard $\partial\Omega$ as  a permeable surface with partial slip, reflecting 
%a non-linear generalization of the Beavers-Joseph problem for Newtonian fluids \citep{Nield09} \kcom{I don't quite understand this sentence... I think just a bit more description of what is meant would make this clearer}.\jcom{I think it was a holdover from a previous version with a different Figure and alternative  definition of $I$. I suppose this whole paragraph could be eliminated if you do not 
%find it useful.} 

Considering the traction of fluid on solid $\vec{t}_I=-(\sigb\vec{n})$ for $\vec{n}$ the interfacial outward normal (pointing into solid), we  set down the following definitions: 
\begin{enumerate}[]
\item A \emph{globally dissipative surface}  $I$  is one for which 
\begin{equation}\label{global}\int_I \vec{t}_I \!\cdot\! \vec{v}\ dS\ge 0, \end{equation} 
\item  a \emph{locally dissipative surface}  $I$  is one for which
\begin{equation}\label{local}\vec{t}_I(\vec{x}) \!\cdot\! \vec{v}(\vec{x}) \ge 0, \:\:
\forall \:\:\vec{x}\in I, \end{equation}
\item and a \emph{strongly dissipative surface} $I$ is locally dissipative surface described by a dissipative surface potential $\psi_I(\vec{v}_I; \vec{x})$, such that
\begin{equation}\label{surf}
\vec{t}_I(\vec{x})=\partial_{\vec{v}}\psi_I(\vec{v};\vec{x}), \:\:\mbox{with}\:\: \vec{t}_I(\vec{x}) \!\cdot\! \vec{v}(\vec{x})\ge0 \ \ \ \text{for all} \ \ \ \vec{x}\in I 
\end{equation} 
\end{enumerate}

The relation (\ref{global}) can be expected to apply on the surface $I$
of a porous medium  with internal  viscoplastic flow in pores whose walls are locally dissipative,
 whereas (\ref{local}) and (\ref{surf}) may fail to apply because 
of the local power input to the fluid (``pV" work ) associated with outflow from $I$.

Owing to the  impermeability of the solid interface, $\vec{t}_I$ can be replaced by the vector of  shear traction $\taub_I=-(\vec{1}-\vec{n}\otimes\vec{n})\sigb\vec{n}$ which provides the power associated with surface slip.  Letting $\varphi_I$ represent the dual potential to $\psi_I$,  we note for example that $\varphi_I(\taub_I)=\ell |\taub|^2/2\eta$ gives the {\it Navier} slip relation 
%\jcom{Ken, why introduce a fictitious shear rate here when a velocity will do? $\vec{v}:=\ell \gdotb\partial_{\taub_I}\varphi_I=\ell \taub_I/\eta$, for a fictitious vectorial shear rate $\gdotb$}
 for the slip velocity $\vec{v}=\partial_{\taub_I}\varphi_I=\ell \taub_I/\eta$ in a Newtonian fluid with viscosity $\eta$ and slip-length $\ell$, for which the the no-slip condition is given by $\ell\equiv 0$.  As an extension of  Navier slip, the strongly dissipative slip relation proposed here describes a wide range of wall-slip phenomena, including nonlinear, anisotropic and spatially inhomogeneous slip, with inhomogeneity represented by dependence on $\vec{x}\in  I $. 

For example, when the anisotropy is determined by a symmetric 
%$2\times2$ tensor $\boldsymbol{\Lambda}$\jcom{why not say "
  surface tensor $\boldsymbol{\Lambda} : \mathcal{T}_I \to \mathcal{T}_I$, transforming vectors in the tangent space $\mathcal{T}_I$ of $I$
% "} 
 we may take $\varphi_I(\taub)$ to be a function of the joint isotropic invariants of $\taub, \boldsymbol{\Lambda}$: \[ |\taub |, \operatorname{tr}\boldsymbol{\Lambda}, \operatorname{tr}\boldsymbol{\Lambda}^2,  \taub \!\cdot\!\boldsymbol{\Lambda}\taub, \ldots\]
an instructive special case being  
\begin{equation}\label{nonlinhom} \varphi_I(\taub;\vec{x})=f(\taub \!\cdot\!\boldsymbol{\Lambda}\taub), \:\:\mbox{with}\:\:\vec{v}_I= 2f'(\taub_I \!\cdot\!\boldsymbol{\Lambda}\taub_I)\boldsymbol{\Lambda}\taub_I, \end{equation} 
where the prime denotes a derivative, and $\boldsymbol{\Lambda}=\boldsymbol{\Lambda}(\vec{x})$ is allowed depend on position $\vec{x}$ on $I$. When $f(s)$ is given by a power law in $s$ one obtains an anisotropic power-law for $\vec{v}_I$ in terms of $\taub_I$. Were we to consider  a permeable surface $I$, then
$\boldsymbol{\Lambda}$ would have to be replaced by a more general linear transformation $\mathbb{R}^3\to\mathbb{R}^3$. 
%\kcomss{I was actually referring to impermeable surfaces as well as permeable ones.  Let me explain: I define the shear stress traction vector $\taub_I$ by $\taub_I=(\vec{1}-\vec{n}\otimes\vec{n})(-\sigb\vec{n})\in \mathbb{R}^3$; an honest  3-vector even though we know it is tangent to a given surface with normal $\vec{n}$.  Under this definition, since $\vec{A}$ acts on $\taub_I$, one should also embed $\vec{A}$ in 3-space and treat it as a $3\times3$ tensor.}\jcom{ Ken, If you wish to extend the definition to permeable surfaces that is o.k. by me. Otherwise, in the (dishonest?) sense of the differential 
%geometry, $\taub_I$  is a two-vector lying in the 2D tangent space
%to a 2D manifold. As I think you implied earlier, all we need for $\vec{A}$
%in that case is the 2x2 matrix representing linear transformations on that tangent space, i.e.
%its "restriction" to the tangent space.  In that case,  $\vec{A}$ must have the 
%form $(\vec{1}-\vec{n}\otimes\vec{n})\bolsymbol{\Lambda}(\vec{1}-\vec{n}\otimes\vec{n})$,  and it suffices to state that $\boldsymbol{\Lambda}$ is a two-tensor. Anyway, I will accept any statement you wish to make on this. }
%}
\section*{\normalsize From locally to globally strong dissipation}
The  local considerations above have immediate implications for the 
existence of systems that are strongly dissipative in a global sense.

Suppose that a strongly dissipative fluid occupies a  region $\Omega_F$ discussed above.   A flow field 
$\vec{v}(\vec{x})$ in $\Omega_F$  is said to be admissible, if $\nabla \!\cdot\!\vec{v}=\vec{0}\!\cdot\!$ and  its associated stress field satisfies 
$\nabla \!\cdot\!\sigb(\vec{x})=\vec{0}$ for $\forall \vec{x}\in \Omega_F$. 
%As a further definition, we call a vector field $\vec{w}(\vec{x}) : F\to\mathbb{R}^3$ globally solenoidal on $\Omega_F$  if $\int_{F}\vec{w}\!\cdot\! \vec{n}\ dS=0$.  
We next suppose that   the fluid exhibits a strongly dissipative slip on $ I $ and    shall then prove that \emph{all admissible  solutions for the flow in $\Omega_F$ must satisfy strongly dissipative boundary conditions  on $F$.}
%
%\kcomss{As you will see, I made a number of changes to the presentation below here, so that all definitions of mathematical things are done first, and then we state our theorems after.  I felt this made the most sense and made it easier to read. I chose to use generic variables in the definitions so as not to confuse with named variables already being used.}

As a general convention, we denote vector-valued functionals  by arguments enclosed in brackets  $[\quad]$, e.g. $
\vec{f}=\vec{f}[\vec{w}]$ denotes a map $\mathfrak{F}\to \mathbb{R}^n$ from the vector field $\vec{w}(\vec{x})$ to the real vector $\vec{f}$, where $\vec{w}(\vec{x})$ is an element  of function space $\mathfrak{F}$ , e.g.~a Banach space.  We 
denote maps $\mathfrak{F}\to\mathfrak{F}$ of one vector field to another  by curly brackets $\{\quad\}= [\quad](\quad)$, i.e. $\vec{f}=\vec{f}\{\vec{w}\}=\vec{f}[\vec{w}](\vec{x})$ denotes a map from the vector field $\vec{w}(\vec{x})$ to the vector field $\vec{f}(\vec{x})$.  Further, for a given vector-valued functional $\vec{f}[\vec{w}]$, we make use of the \emph {Fr\'echet (or variational) derivative} of $\vec{f}[\vec{w}]$, denoted
 $\delta_{\vec{w}}\vec{f}\{\vec{w}\}$, which is the mapping $\mathfrak{F}\to\mathfrak{F}$ that satisfies 
\begin{equation}\label{frech}
\langle \ \delta_{\vec{w}}\vec{f}\{\vec{w}\}\ , \ \vec{h}\ \rangle \equiv \int_{\mathcal{R}}\vec{h}(\vec{x}) \!\cdot\!\delta_{\vec{w}}\vec{f}[\vec{w}](\vec{x}) \ d\mathcal{R}(\vec{x}) =\lim_{\epsilon\rightarrow 0}\frac {\vec{f}[\vec{w}+\epsilon\vec{h}]-\vec{f}[\vec{w}]}{\epsilon}  \end{equation}
for all test functions $\vec{h}\in \mathfrak{F}$, where $\mathcal{R}$ is the domain of the vector field $\vec{w}(\vec{x})$. The bilinear pairing above defines a functional of $\vec{h}$ referred to as the \emph{Fr\'echet differential} denoted $ \delta_{\vec{w}}\vec{f}[\vec{h}]=\langle \ \delta_{\vec{w}}\vec{f}\{\vec{w}\}, \ \vec{h}\ \rangle$.
%\footnote{To clarify the independent variable, we include an additional space before the argument of a function whenever the function is being expressed as an operation on another function with a different codomain.}

Our main results pertain to the nature of the mapping from boundary velocity fields $\vec{v}_F(\vec{x}): F \to \mathbb{R}^3$ to the corresponding boundary traction field $\vec{t}_F(\vec{x}): F \to \mathbb{R}^3$, i.e. the mapping $\vec{t}_F=\vec{t}_F\{\vec{v}_F\}$.  To be specific, $\vec{t}_F=\sigb\vec{n}$ is the surface traction field on $F$ that arises from the admissible bulk flow that has  $\vec{v}_F$ as boundary condition.  Due to incompressibility, the set of boundary velocites is further restricted to be solenoidal, i.e. $\int_{F} \vec{v}_F\cdot \vec{n}\ dS=0$, where subscript $F$ denotes a function whose domain is restricted to the surface $F$.  
%
%As with all incompressible flows, the corresponding stress field is unique only up to an additive  hydrostatic pressure.  To choose one representative traction field, we suppose the extra pressure is selected such that $\vec{t}_F$ is solenoidal over $F$, i.e. that \crd \[\int_F \vec{t}_F\cdot\vec{n} \ dS=0. \] \cbk 

We shall show that 
\begin{equation}\label{strict}
\int_{F}\vec{t}_F\{\vec{v}_F\} \!\cdot\!\vec{v}_F\ dS\ge0\ ,
\end{equation}
 and secondly, that there exists a functional $\Psi[\vec{v}_F]$ such that
\begin{equation}\label{strong} 
\delta_{\vec{v}_F}\Psi\{\vec{v}_F\}= \vec{t}_F\{\vec{v}_F\}.
\end{equation}
which establishes a direct analogy to  (\ref{const}).  Note that $\vec{t}_F\{\vec{v}_F\}$ is defined up to the addition of an inessential uniform surface pressure. \\
%\jcom{Ken, I hope this suffices
%to make the point that concerned you.} \\ 

\noindent \emph{Proof:}
\\
Assuming sufficiently smooth potentials and fields,  we may express 
%\kcomss{Just tweaked a couple words in this next sentence... now I am comfortable!}
the quasi-static stress balance in terms of {\it virtual power} as 
\begin{equation}\label{virtual}  \int_F\vec{t}_F\{\vec{v}_F\} \!\cdot\!\vec{u}_F\ dS =\int_I\vec{t}_I(\vec{v}) \!\cdot\!\vec{u}_I\ dS+\int_{\Omega_F} \vec{S}(\vec{E}):\nabla\vec{u}\ dV,  \end{equation}
for any admissible velocity field $\vec{v}(\vec{x})$, with $\vec{E}=$ sym $\nabla\vec{v}$, where $\vec{u}(\vec{x})$ is any solenoidal ``test field". 
%For the present purposes, we shall 
%allow singular test fields, with discontinuities of the type discussed below.  \kcom{Do we still need to worry about mentioning singular test fields here, now that we have our new porous flow explanation, which does not necessitate considering flow with jumps?  We can always bring it up, as we do, when we get to Eq 21.}

Now,  we assume here and in the following that the choice of boundary field $\vec{v}_F(\vec{x}), \:\:\vec{x}\in F,$ uniquely determines the bulk flow $\vec{v}(\vec{x})$ and strain-rate $\vec{E}(\vec{x})$. 
 In the case of multiplicity, a possibility we do not entertain here, one could presumably restrict the solution to a particular solution branch. 
 
 In the case of an impermeable surface, $\vec{t}_I$ can be replaced by $\taub_I$ above,  and, invoking locally strong 
dissipation in $\Omega_F\cup I$, we have the local relations \begin{equation}\label{loc}\vec{S}=\partial_{\vec{E}}\psi(\vec{E}; \vec{x}), \:\:\taub_I=\partial_{\vec{v}}\psi_I(\vec{v}; \vec{x}),   \end{equation} 
where $v_n=\vec{v}\!\cdot\! \vec{n}=0$ on $I$. We have allowed the potential $\psi$ to depend on position $\vec{x}$ to reflect a possible dependence on inhomogeneous and/or evolutionary  structure tensors of the type discussed above.  Note that time variations in velocity  do not imply instantaneous variations in these tensors, which are presumably governed by smooth ODEs of the form 
(\ref{fabric}). 

The preceding relations yield a linear functional of the  test velocity $\vec{u}_F$ on $F$:
\begin{equation}\label{strong1} L_{\vec{v}_F}[\vec{u}_F]:= \int_F\vec{t}_F\{\vec{v}_F\} \!\cdot\!\vec{u}_F\ dS=
\int_{\Omega_F}  \partial_{\vec{E}}\psi(\vec{E}; \vec{x}):\nabla\vec{u}\ dV 
+\int_I \partial_{ \vec{v}}\psi_I( \vec{v}; \vec{x}) \!\cdot\!\vec{u}_I\ dS
\end{equation} 
 It is evident  that
in the case where $\vec{u}\equiv\vec{v}$ in $\Omega_F$ that the dissipation of $\vec{v}(\vec{x})$ is given by  
 \begin{equation} \mathfrak{D}[\vec{v}]=L_{\vec{v}_F}[\vec{v}_F]=\int_F \vec{t}_F\{\vec{v}_F\}  \!\cdot\!\vec{v}_F\ dS=\int_{\Omega_F}  \partial_{\vec{E}}\psi(\vec{E}; \vec{x}):\vec{E}\ dV 
+\int_I \partial_{ \vec{v} }\psi_I( \vec{v}_I; \vec{x}) \!\cdot\!\vec{v}_I\ dS
\ge 0\label{diss}\end{equation}
 as all integrands in the last expression are necessarily positive by local strong dissipation.

 By employing  (\ref{frech}), it can also be easily proved  that:\begin{equation}\label{diss2} L_{\vec{v}_F}[\vec{u}_F]=\langle \delta_{\vec{v}_F}\Psi, \ \vec{u}_F\rangle, \:\:\mbox{where}\:\: \Psi[\vec{v}_F]= \int_{\Omega_F}\psi(\vec{E}; \vec{x})\ dV + \int_{I }\psi_I(\vec{v};\vec{x})\ dS \end{equation}
the latter relation following from the fact that $\vec{v}(\vec{x})$ and $\vec{E}(\vec{x})$ are determined by $\vec{v}_F$. The above, together with (\ref{strong1}) implies the desired result
\begin{equation}
\delta_{\vec{v}_F}\Psi\{\vec{v}_F\}= \vec{t}_F\{\vec{v}_F\}.
\end{equation}
In the Appendix, we indicate that the above results follow from a tentative 
extension of Edelen's formula for finite dimensional vector spaces to infinite 
dimensional (e.g. Banach) function spaces.\\
\\
\noindent \emph{Remarks:}  
\begin{enumerate}[i.]
\item In the dual description with interfacial boundary condition $\vec{v}_I=\partial_{\vec{\taub}_I}\phi_I(\vec{\taub}_I)$ on $I$, we can swap variables and show  that  dual functional $\Phi[\vec{t}]$ exists, mapping traction fields $
\vec{t}$  on $F$ to a scalar, such that 
\begin{equation}\label{dual2}
\delta_{ \vec{t}_F }\Phi \{ \vec{t}_F \}= \vec{v}_F\{ \vec{t}_F \}.
%\delta_{\vec{t}}\Phi\ [\vec{p}] =\int_{F}\vec{v}_F\cdot\vec{p}\ dS
\end{equation}
%for all globally solenoidal test fields $\vec{p}$ on $F$, where $\vec{p}$ now represents a traction. 

\item The relation (\ref{strong}),  or its dual  (\ref{dual2}),   confirms the perhaps intuitively obvious fact that the 
dissipation and dissipation potential are given directly by  the boundary field $\vec{v}_F$ on $F$.   To compute the corresponding boundary traction $ \vec{t}_F$ in full, one would need to compute the actual fluid velocity field  $\vec{v}(\vec{x})=\vec{v}\{\vec{v}_F\}$ in $\Omega_F$.  However, the fact that the boundary field $ \vec{t}_F$ must arise as a Fr\'echet  derivative places a restriction on $\vec{t}_F$ that can be exploited without specification of the full flow field.
 Based on simplified boundary conditions, the following sections will make use of  this derivative in a form appropriate  to finite dimensional vector spaces.

 \item In the case of unbounded regions, we should replace  integrals like those
 in  (\ref{diss})-(\ref{diss2})    by appropriate averages over 
 $\Omega_F, I$ and $F$, which we do not bother to define explicitly.

\item There are  direct analogies between the theorem proven above and the global theorems of hyperelasticity, owing to the fact that both are based on constitutive potentials.  
We note however, that the non-trivial slip boundary conditions in this treatment would correspond to an unconventional {\it Robin}  
%unusual \jcom{"somewhat unconventional Robin-type" instead of "unusual"?} 
traction-displacement condition in the elastic analogy. 
As in hyperelasticity, the admissible solution minimizes the functional for  appropriate boundary conditions.  Unlike hyperelasticity, where the functional  represents total potential energy, the functional $\Psi$ provides in fact a \emph{lower bound} for total dissipation, with equality in the case of a linear rheology/slip condition, or  proportionality in the case of homogeneous  potentials.  The lower-bound property is easily proved based on the guaranteed convexity of the underlying local dissipation potentials.

\item The proof above relies on  $C^1$   smoothness of the underlying potentials, such that a unique derivative and therefore a unique stress can be assumed in Eq \ref{const}.  However, we note that certain dissipation potentials $\varphi$ may become ``kinked",  with loss of convexity,   such as those representing the above-mentioned yield-stress fluids.   Although the stress may fail to be uniquely defined at zero strain rate, we shall assume that the solution 
to the variational problem, including the spatial location of  the associated yield surfaces, can be rendered unique by various techniques,  such as those employed to determine ``singular minimizers" for non-convex hyperelasticity \citep{Ball01}. In this respect, we note that existing treatments \citep{Denn11} of yield-stress fluid appear to overlook the associated extremum principles and the applicability of variational methods.

\item For the sake of definiteness, we have adopted in  (\ref{virtual}) a relatively strong form of virtual power. However,  certain fluids may exhibit material instability, arising from  a  loss of convexity of the dissipation potential and leading to  the formation of singular surfaces.  At such surfaces  the  velocity field may become non-differentiable and even discontinuous, e.g. across an infinitely thin ``shear band".  Indeed, the assumed slip on the surface $I$ has this same character. To cover such singular behavior, we may express (\ref{virtual}) in the weaker form involving jumps $[[\vec{u}]]_J$ in $\vec{u}$ on a discrete
set of singular surfaces $J$ across which the traction $\vec{t}_F$ is continuous:
\begin{equation}\label{weak} \int_F\vec{t}_F\{\vec{v}_F\} \!\cdot\!\vec{u} dS =\sum_J \int_J \vec{t}_J \!\cdot\![[\vec{u}]]_J\ dS+\int_{\Omega_F} \vec{S}(\vec{E}):\nabla\vec{u}\ dV, \:\: \mbox{where} \:\: \vec{t}_J=\left.\sigb \!\cdot\!\vec{n}\right\vert_J \end{equation}
 In this case,  we admit discontinuous test velocity fields, and  (\ref{virtual}) is a special case in which $\vec{u}={\bf 0}$ on $I$, with jump $[[\vec{u}]]= \vec{u}_I$, and is otherwise continuous.  
 \end{enumerate}
We now illustrate the utility of the above results by the application to 
two physically interesting problems involving flow around  
a geometrically complex solid surface.
  
\section*{\normalsize  Generalized Darcy flow in porous media}

For the  application to viscoplastic flow in a porous medium, we apply the above     to two types of boundary condition on the fluid surface $F$ of a given porous body, namely  prescribed velocity and  prescribed stress.  

\subsubsection*{\it Traction BC}
We first consider  the  boundary condition
\begin{equation}
\vec{t}_F =-(\vec{g} \!\cdot\!\vec{x})\vec{n}+\vec{T}\vec{n}, 
\:\:\vec{x}\in F,  \label{stressBC}
\end{equation}
for a constant vector $\vec{g}$ and constant tensor $\vec{T}$.  Above, $\vec{g}$ represents a pressure gradient applied to the fluid boundary of the domain and $\vec{T}$ is a symmetric boundary stress tensor. 

As in the preceding subsection, we assume now that the boundary condition (\ref{stressBC}) determines the flow in $\Omega_F$.  The  
functionals of $\vec{t}_F$ in (\ref{dual2}) now reduce to  
functions of $\vec{g}$ and $\vec{T}$,  with dissipation and potentials obeying
\begin{equation}\begin{split}\label{flow}& \mathcal{D}[\vec{t}_F]=D(\vec{g}, \vec{T})=V_F(\vec{g}\!\cdot\!\overline{\vec{v}}+ \vec{T}\!:\!\boldsymbol{\overline{\vec{D}}}), \:\: \mbox{with}
\\&\overline{\vec{v}}(\vec{g}, \vec{T})=\frac{1}{V_F}\int_{\Omega_F}\vec{v}\ dV=\partial_{\vec{g}}\Phi(\vec{g}, \vec{T}), 
\\&\boldsymbol{\overline{\vec{D}}}(\vec{g}, \vec{T})=\frac{1}{V_F}\text{sym}\left\{\int_{\Omega_F}\nabla\vec{v}\ dV+\int_I \vec{n}_I\otimes\vec{v}_I\ dS\right\}=\partial_{\vec{T}}\Phi(\vec{g}, \vec{T}),\\&\mbox{and}\:\: \Phi(\vec{g}, \vec{T})=\left.\Phi[\vec{t}_F(\vec{x})]\right\vert_{\vec{t}_F(\vec{x})=-(\vec{g} \cdot\!\vec{x}) \vec{n}+\vec{T}\vec{n}}
\end{split} \end{equation}
where $\vec{n}_I$ is the inner normal to the surface $I$.  
The second and third lines of  (\ref{flow}) follow from the vanishing of  $v_n=\vec{v} \!\cdot\!\vec{n}$ on $I$, and the relations
\begin{equation}\begin{split}&\int_F \vec{x}v_n \:dS=\int_{\partial\Omega_F}\vec{x}v_n \:dS=
\int_{\Omega_F} \nabla\!\cdot\!(\vec{v}\otimes\vec{x})\: dV=\int_{\Omega_F}\vec{v}\:  dV,\\&\mbox{and}\:\:\int_F\vec{n}\otimes\vec{v}\ dS=\int_{\partial\Omega_F}\vec{n}\otimes\vec{v}\ dS-\int_I\vec{n}\otimes\vec{v}\ dS=\int_{\Omega_F}\nabla\vec{v}\ dV+\int_I \vec{n}_I\otimes\vec{v}_I\ dS \label{avgs}\end{split}\end{equation}
These results establish the volume average velocity, $\bar{\vec{v}}$, and deformation rate, $\bar{\vec{D}}$, as respective conjugates of $\vec{g}$ and $\vec{T}$.  Note that we could have also obtained the result for average deformation gradient by considering 
a singular test field with  $\vec{u}=\vec{0}$ and $[[\vec{u}]]=\vec{u}_I$ on $I$, as subsumed by (\ref{weak}).  The fact that the drag laws for  $\bar{\vec{v}}$ and $\bar{\vec{D}}$ both must emerge from a single potential function $\Phi$ is a significant restriction on the functional forms of each.

\subsubsection*{\it Velocity BC}
We now consider flow induced by an applied  velocity boundary condition
\begin{equation}
\vec{v}(\vec{x})=\vec{q}+\vec{L}\vec{x}, 
\:\:\vec{x}\in F, 
%%\& \:\: [[\vec{u}]](\vec{x})= \vec{u}_I, \:\: 
%\vec{x}\in I 
\label{velBC}
\end{equation}
where $\vec{q}$ is a constant velocity vector and $\vec{L}$ is a constant velocity-gradient tensor.  
The incompressibility constraint on $\vec{v}$ then requires that
\begin{equation} \int_F v_ndS=\vec{a}\!\cdot\!\vec{q} + \vec{A}\!:\!\vec{L}=0, \:\:\mbox{where}\:\: 
\vec{a}= \int_F \vec{n}dS \:\:\mbox{and}\:\: \vec{A}=\int_F \vec{n}\otimes\vec{x} \ dS\label{incomp}\end{equation}
The vector $\vec{a}$ and the second rank tensor $\vec{A}$ represent an ostensibly new class of 
structure tensors for a representative volume element (RVE), given generally by the $n$th rank moment tensors:
\begin{equation} \boldsymbol{\mathcal{A}}^{(m)}= \int_F \vec{n} (\otimes\vec{x})^{m-1} dS, \:\: m=1, 2,\ldots\label{struct}\end{equation} 
 The lowest moment $\vec{a}$ equals the vectorial excess of out-flow over in-flow area, and $\vec{A}$  can be regarded as  a second-rank  ``fabric" tensor, by loose analogy to a tensor  employed to describe the anisotropy of granular media. It is clear that the 
 overall dissipation potential must eventually be given as a function $\Psi(\vec{q},\vec{L}, \vec{a}, \vec{A})$, subject to the restriction (\ref{incomp}).
 
 %\subsubsection*{\it Isotropic media}
We now specialise to isotropic  media, defining an isotropic medium of {\it degree} $M$  as one for which the structure tensors of order $m=1, 2, \ldots, M$ are 
 invariant under the transformation $\vec{n}, \vec{x} \rightarrow
 \vec{Qn}, \vec{Qx}$, where $\vec{Q}$ is an arbitrary constant orthogonal tensor. 
 This requires that $\boldsymbol{\mathcal{A}}^{(m)}$ be a scalar multiple of the isotropic tensor 
 of order $m$, for $m=1, 2, \ldots, M$ and, hence, that ${\bf a}={\bf 0}$ and $\vec{A}\propto \vec{I}$ for the case $M=2$ considered here. 
 
 Hence, for an isotropic medium of degree 2,  (\ref{incomp}) implies that $\vec{q}$ and $\vec{L}$, with tr\:$\vec{L}=0$,  can be chosen independently.  
 %\jcom{Can we restrict this in some way to cubic symmetry, where we know that the linear
% form of Darcy's law is isotropic?} 
A common example from homogenization theory  is a periodic porous structure, for an RVE consisting of a single periodic cell,  discussed briefly below.

Assuming  that the boundary condition (\ref{velBC}) determines a unique solution to the flow in $\Omega_F$,   the 
functionals of $\vec{v}_F$ in (\ref{diss})  reduce to ordinary 
functions of $\vec{q}$ and $\vec{L}$  with dissipation given by   
\begin{equation}\label{force}\begin{split}& \mathcal{D}[\vec{v}_F]=D(\vec{q}, \vec{L})=\vec{f}\!\cdot\!\vec{q}+ \boldsymbol{\Sigma}\!:\!{\bf L}, \:\: \mbox{with}\:\:
\\
&\vec{f}(\vec{q}, \vec{L})=\int_F\vec{t}_F\{\vec{v}_F\}\ dS=\partial_{\vec{q}}\Psi(\vec{q}, \vec{L}), 
\\
&\boldsymbol{\Sigma}(\vec{q}, \vec{L})=\int_F\vec{t}_F\{\vec{v}_F\}\otimes\vec{x}\ dS=\partial_{\vec{L}}\Psi(\vec{q}, \vec{L}),
\\
& \mbox{and}\:\: \Psi(\vec{q}, \vec{L})=\left.\Psi[\vec{v}_F(\vec{x})]\right\vert_{\vec{v}_F(\vec{x})=\vec{q}+\vec{L}\vec{x}}. 
\end{split}\end{equation} Here $\vec{f}$ is the force and $\boldsymbol{\Sigma}$ the  moment of traction acting on $F$, which can be regarded as the contribution of $F$ to the volume average  stress
\begin{equation}\label{avgstress} \frac{1}{V_F}\int_{\Omega_F}\sigb dV 
= \frac{1}{V_F}\int_{\Omega_F}\rm{div}( \vec{x}\otimes\sigb) dV
= \frac{1}{V_F}\int_{\partial\Omega_F}\vec{t}\otimes
\vec{x}dS\end{equation}
The  relations (\ref{force})-(\ref{avgstress}) provide yet another extension of Darcy's law which, in contrast to the case of stress boundary conditions, 
bears a rather opaque relation to the linear version.  Still, it bears emphasizing that the emergence of $\boldsymbol{\Sigma}(\vec{q},\vec{L})$ and $\vec{f}(\vec{q},\vec{L})$ from a single potential $\psi(\vec{q},\vec{L})$ is a notable restriction on the functional forms of both quantities.

In closing here we note that higher gradient theories would involve higher-order 
  structure tensors of the type  (\ref{struct}). 

\subsubsection*{\it Permeability relations}
The prior subsections apply to  global  flow relations, and we can apply a key result displayed in (\ref{flow}), namely,
\begin{equation}\label{pressure_grad_BC}
\bar{\vec{v}}=\partial_{\vec{g}}\Phi, \ \ \ 
\end{equation}
to describe the average or {\it homogenized} permeability of a non-Newtonian fluid within a regular porous solid,  with  surface slip between  fluid and solid that may  be both  non-linear and  non-uniform. 

As one example of  a structure with a single well-defined permeability, consider an  
idealized periodic RVE,  composed of a repeated tiling of box-shaped porous elements.  Owing to  periodicity, the permeability of a large sample of such a material can be defined by treating the flow through one such element  induced by an applied pressure  on the element of the form (\ref{stressBC}), with $\vec{T}=\vec{0}$.  The pressure gradient $\vec{g}$ provides an arbitrary pressure difference between parallel faces of the cell, and (\ref{pressure_grad_BC}) shows that mean fluid velocity $\bar{\vec{v}}$ and the applied pressure gradient, $\vec{g}$ are connected through the derivative of a potential $\Phi=\Phi(\vec{g})$.

Following a previous mathematical analysis  \citep{Goddard14},  the preceding relationship (\ref{pressure_grad_BC}) can also be expressed in a \emph{pseudo-linear} form in terms of a permeability tensor $\vec{K}=\vec{K}(\vec{g})$ defined by
\begin{equation}\label{perm}
\bar{\vec{v}}=\vec{K}(\vec{g})\ \vec{g}.
\end{equation} 
where $\vec{K}$ must take the form of the Hessian $\partial_{\vec{g}\vec{g}}^2\chi$ of a complimentary  function $\chi(\vec{g})$  derived from $\Phi(\vec{g})$ \citep{Goddard14}.  
This in turn implies the following symmetry and differetial compatibility conditions on the permeability tensor:
\begin{align}\label{porous_compat}
\vec{K}(\vec{g})=\vec{K}(\vec{g})^T \ \ \ \  \text{and} \ \ \ \deriv{K_{ij}}{g_k}=\deriv{K_{jk}}{g_i}
\end{align}
for all $i, j, k\in\{1,2,3\}$.  

Thus, by means of  (\ref{perm})  we extend  the standard linear (Onsager) symmetry to the flow of rheologically non-linear fluids through porous solids  with non-uniform and nonlinear interfacial slip.  In contrast to  the linear case, $\vec{K}$  depends on $\vec{q}$, but as   with linear case $\vec{K}$ is non-negative definite,  reflecting  non-negativity of dissipation. The constraints imposed by (\ref{porous_compat}) represent a severe restriction on the form of $\vec{K}(\vec{q})$, as a substantial extension of the linear theory.

 \section*{\normalsize Nonlinear mobility in effective-slip problems}
As a generalization of the corresponding Stokes-flow problem \citep{Kamrin11}, 
we consider the far-field condition on the top surface $F$ of a viscoplastic shear flow bounded below by a textured surface solid surface $I$ on which strongly dissipative slip may occur. The interface $I$ represents an impermeable solid with possibly non-uniform partial slip distribution.
%
%\kcom{We should adjust this figure so that: (1) the bulk velocity field is displayed as $\vec{v}$ rather than $\vec{u}$ for notational consistency,  (2) the top surface $F$ should intersect the $x_3$-axis at a position denoted $z_H$ to be consistent with the text, and (3) the $\taub$ vector should agree locally with the gradient of the flow field at $z_H$.  This third point is a bit nit-picky, I know...}\jcom{(1) o.k. (2) o.k. but I am uncomfortable
%with the idea that we have an actual lid on the flow, when I would have thought we were 
%dealing with an asymptotic relation. Hence, I would not have used Eq. (37) as you do. (3) Why would we expect these to be parallel for 
%a non-linear anisotropic problem. Even for the linear case they are not generally parallel, no?}\kcom{Regarding (2) I see your point, though I think the derivation is much easier if we choose some generic $z_H$ that lives in the asymptotic range --- by choosing a $z_H$ the picture conforms more closely to the circumstances displayed in our main theorem. (3)  Ah, ok.  I hadn't thought about anisotropic material laws.  Sure, that could be possible then. When the constitutive law is isotropic the direction of simple shearing and that of simple shear stress must align, but this is not so in the anisotropic case.}
 \begin{figure}[h]
\begin{center}
\includegraphics[width=3in]{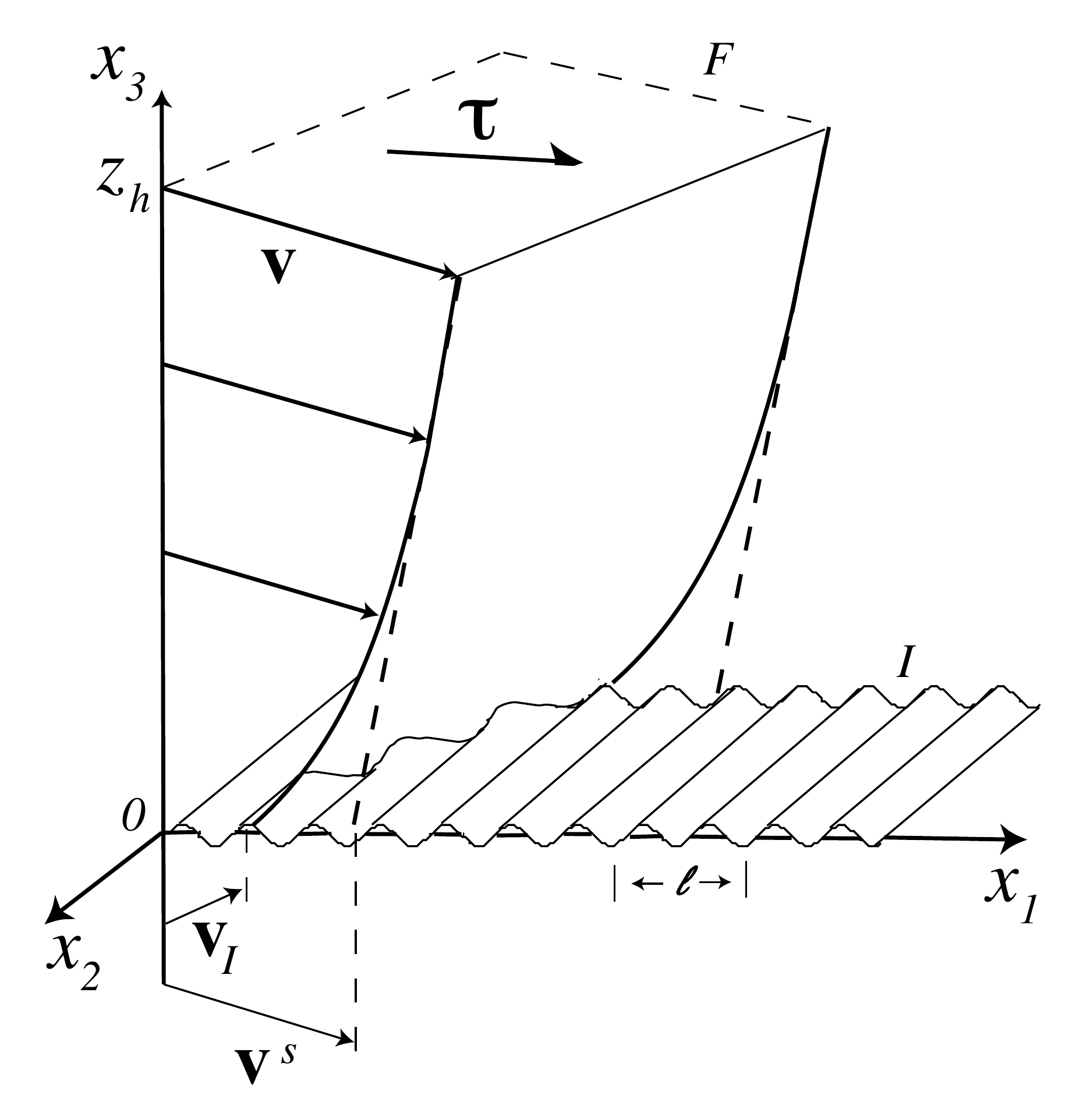}
\caption{Effective slip above a textured surface. 
% \kcom{I see you switched the $\vec{u}$ to a $\vec{v}$ which is good, but it looks like you also switched $\vec{u}^s$ to $\vec{v}^s$.  Because the effective slip velocity is always written (at least in previous literature) as $\vec{u}^s$, I had kept with that notation in this section. Mind switching that one back?  Sorry this is confusing!}\jcom{Ken, pardon me if I turn the tables on you - who wanted to 
%eliminate the confusion between test velocity $\vec{u}$ and actual velocity  $\vec{v}$. Unless I am mistaken, $\vec{v}^s$ represents the latter, no? Why should we be constrained
%by the notation of previous literature?}\kcom{Tables successfully turned! I suppose there's no good reason to keep with the older denotation.  I have switched $\vec{u}^s$ to $\vec{v}^s$ throughout.}
}
\label{Fig:flow}
\end{center}
\end{figure}

 Adopting Cartesian coordinates $x_i$ with unit basis $\vec{e}_i$ for  $i=1,2,3,$ we consider the case of  a surface $I$ of infinite extent having mean elevation  $x_3=0$. As illustrated in Fig.~\ref{Fig:flow}, the layer is bounded  above by a flat plane $F$  situated at an arbitrary position $x_3=z_H$,  and driven by a uniform parallel shearing traction $\taub,$ with $\taub\cdot \vec{e}_3=0$.   We assume the surface texture is a repeating pattern, and hence the induced flow field is periodic in  horizontal planes.  That said, we place no constraints on the period length so this could still represent an arbitrarily large surface pattern.

In the absence of external body forces or overall pressure gradient, and  given $x_3\gg \ell $, where $\ell$ is a characteristic length scale related to the surface texture,  we assume that an admissible velocity field will exhibit  far field behavior consisting of  uniform simple shear plus an apparent slip relative to the surface $I$:
\begin{equation}\label{far} \vec{v}(x_1,x_2,x_3\gg\ell)= \gdotb\ x_3+ \vec{v}^s, \:\:\mbox{with}\:\:\gdotb\!\cdot\!\vec{e}_3=0, \:\: \vec{v}^s\cdot \vec{e}_3=0, \:\:\vec{E}=\mbox{sym} (\boldsymbol{\dot{\gamma}}\otimes \vec{e}_3), \end{equation} 
where the  shear rate vector, $\gdotb$, and the effective slip velocity, $\vec{v}^s$, are assumed to become asymptotically  independent of position $x_2, x_3$ on $F$. 

Given the fluid rheology, the surface pattern, and the slip relation on  $I$, the problem  is to find the relation  between $\vec{v}^s$ and $\taub$ on $F$.  This relation may  be taken to represent an effective slip boundary condition for the large-scale description of the flow on length scales $\gg \ell$.  
In the case of a Newtonian  fluid with linear slip on $I$, it has been shown \citep{Kamrin11} that there exists a positive symmetric  $2\times2$ \emph{surface (Onsager) mobility tensor} $\vec{M}=[M_{ij}]$,  such that: 
\begin{equation}\label{mobility}
\vec{v}^s=\vec{M}\ \taub , \:\:\mbox{or} \:\: v^s_i =M_{ij}\tau_j, \:\: i,j=1,2
\end{equation}
on a Cartesian basis. Here, we extend the analysis to nonlinear viscoplastic fluids  with strongly dissipative nonlinear inerfacial slip.

As as a helpful albeit not essential device, we define our domain $\Omega$  as a cell $[-\ell_1,\ell_1]\times[-\ell_2,\ell_2]\times[z_L,z_H]$, where $z_L$ lies within the rigid solid beneath $I$ and $\ell_1$ and $\ell_2$ are the respective period lengths in the horizontal plane. Owing to the flow periodicity, the vertical walls need not be considered part of $\partial\Omega$,  and we have $F=[-\ell_1,\ell_1]\times[-\ell_2,\ell_2]\times z_H$ and $S=[-\ell_1,\ell_1]\times[-\ell_2,\ell_2]\times z_L$. Then, the formulae in  (\ref{force}) for constant velocity  $\vec{q}$ on $F$ are applicable,
and one readily finds, upon choosing $\vec{q}$ to have a vanishing $x_3$ component, that 
\begin{equation}\label{tangent}  D = \taub\cdot\vec{q}=\taub\cdot\gdotb\ x_3 + \taub\cdot
\vec{v}^s, \ \ \ \  \taub = \partial_{\vec{q}}\Psi(\vec{q}), \ \ \vec{q}=\partial_{\taub}\Phi(\taub).\end{equation}
Here, the surface force $\vec{f}$ in (\ref{force}) is  $\taub A_F$, and one absorbs $A_F=4\ell_2\ell_2$ into the potential.  The last relation above follows as the Legendre convex conjugate of the prior relation, and we obtain the same potential $\Phi$ from the Legendre conjugate as the one described in (\ref{force}).
Because the fluid rheology is strongly dissipative, with  $\vec{D}(\vec{S})=\partial_{\vec{S}}{\phi}$ for some given dissipation potential $\phi$, the flow at the top boundary satisfies \begin{equation}
\gdotb(\taub)=\deriv{\phi^H}{\taub}(\taub) \ \ \ \text{for} \ \ \ \phi^H(\taub)\equiv2\ \phi(\taub\otimes\vec{e}_3+\vec{e}_3\otimes\taub).
\end{equation}
Hence, by the definition of effective slip, we may write
\begin{equation}\label{slip2}
\vec{q}=z_H\deriv{\phi^H}{\taub}(\taub) +\vec{v}^s.
\end{equation}
Combining (\ref{tangent}) and (\ref{slip2}), and defining $\phi^s(\taub)=\Phi(\taub)-z^H\phi^H(\taub)$, we arrive at
\begin{equation}
\vec{v}^s=\deriv{\phi^s}{\taub}.
\end{equation}
We thus find that the slip velocity must arise from the shear traction as the gradient of a \emph{slip potential}.  
As shown in a previous analysis \citep{Goddard14} 
this implies the existence of a function $\chi^M$ such that the pseudo-linear form
\begin{equation}
\vec{v}^s=\vec{M}(\taub)\ \taub
\end{equation}
arises from a mobility obeying $\vec{M}(\taub)= \partial^2\chi^M(\taub)/\partial\taub^2$.  Accordingly the mobility must obey the following symmetry and differential compatibility conditions:
\begin{align}\label{mob_compat}
\vec{M}(\taub)=\vec{M}(\taub)^T \ \ \ \  \text{and} \ \ \ \deriv{M_{ij}}{\tau_k}=\deriv{M_{jk}}{\tau_i}
\end{align}
for all $i, j, k\in\{1,2\}$.  In the linear case, the constant $\vec{M}$ tensor can be  shown to be symmetric via Lorentz reciprocity \citep{Kamrin11}, whereas (\ref{mob_compat}) establishes the symmetry of the  mobility tensor  for the more general case of any strongly dissipative rheology and interfacial slip relation.

\section*{\normalsize Conclusions}
The foregoing analysis addresses a general class of problems involving viscoplastic flow adjacent to solid boundaries with nonlinear and possibly nonuniform slip conditions.  When the bulk rheology  and interfacial slip take on strongly dissipative forms, the relationship between the  velocity  and traction on fluid boundaries are interrelated by global variational derivatives.  Exploiting  this fact, we have established symmetries of viscoplastic drag laws for two illustrative applications. 

In the first example, involving flow in porous media, the  global conjugacy  between traction and velocity on the external fluid boundary is employed to determine permeability relations for strongly dissipative fluids exhibiting slip along the pore walls. The  permeability is shown to be characterized by a symmetric  positive-definite  tensor that depends on the flow,  with partial derivatives  satisfying a set of compatibility conditions. 

In the second example, we have explored the implications of  strong  dissipation for viscoplastic flow over textured surfaces.  As an extension of the linear case, we find once  again a symmetric positive-definite mobility tensor which connects the far-field  shear traction to the apparent slip velocity and satisfies once more a set of differential compatibility conditions. 

While the foregoing  analysis has focused on impenetrable solid boundaries and incompressible fluids, many of the results may remain qualitatively correct  whenever these conditions are relaxed. In particular, note that on replacing $\vec{S}$ by $\sigb$ and $\vec{E}$ by $\vec{D}$ one can obtain a theory for viscoplastic flow with variable density, such as might occur in a dilatant granular media or particle suspension.  Such variants on the current model may   engender  additional complications that seem worthy of further analysis.

Although we have not explored various consequences of the associated extremum principles for
 strongly dissipative fluids, these  may prove a convenient tool for other applications.
Important examples are the the derivation of continuum models for dispersions of rigid or viscoplastic particles in viscoplastic fluids, or the determination of the spatial configuration of yield surfaces in yield-stress fluids.

\appendix
\section*{\normalsize Appendix: Functional for the dissipation potential}
Edelen's formula \citep{Goddard14} for the dissipation potential $\psi(\vec{v})$ in a finite vector space $\mathbb{R}^n$ is given
in terms of the dissipation $D(\vec{v})$ by 
\begin{equation} \psi(\vec{v})= \int_0^1D(\lambda \vec{v})\lambda^{-1}d\lambda, \end{equation}
with conjugate force given by $\vec{f}=\partial_{\vec{v}}\psi$ in the dual
space. 

Without attempting a rigorous proof here,  we offer as a conjecture a  generalization to a function space (e.g.~Banach space) $\mathfrak{F}$ of vector-valued velocities   $\vec{v(\vec{x})}$ and (dual space) of vector-valued forces $\vec{f}\{\vec{v}\}$,  $\vec{x}\in \mathcal{R}$. It appears that one might achieve this generalization rigorously by extending Edelen's  differential geometric treatment to Banach spaces \citep{Lang99}\footnote{We are indebted to Professor Reuven Segev for this reference.}.  Thus, with dissipation defined by the pairing 
\begin{equation}  \mathcal{D}[\vec{v}]=\langle \vec{v}, \vec{f} \rangle := \int_\mathcal{R}\vec{v}(\vec{x})\cdot \vec{f}\{\vec{v}\}\ d\mathcal{R}:= \int_\mathcal{R}\vec{v}(\vec{x})\cdot \vec{f}[\vec{v}](\vec{x})\ d\mathcal{R},\end{equation}
the corresponding function-space forms for potential and associated force are: 
\begin{equation}\label{funspace} \Psi[\vec{v}] = \int_0^1\mathcal{D}[\lambda\vec{v}]\lambda^{-1}d\lambda, \:\:\mbox{with}\:\: \delta _{\vec{v}}\Psi\ [\vec{q}]=\langle\vec{f}, \vec{q}\rangle=\int_\mathcal{R} \vec{f}\{\vec{v}\}\cdot\vec{q}\ d\mathcal{R}, \:\:\mbox{i.e.}\:\: \vec{f}\{\vec{v}\}=\delta_{\vec{v}}\Psi\ \{\vec{v}\}.\end{equation}
It is not too difficult to verify the final relation in (\ref{funspace}) by making 
use of the properties of Fr\'echet derivatives, whose existence is guaranteed by {\em Moreau's Theorem} for convex functionals \citep{Rockafellar97}.

For example, with solenoidal  velocity field $\vec{v}$ and associated
deviatoric straining $\vec{E}$, we have
\begin{equation} \mathcal{D}[\vec{v}]= \int_\Omega D(\vec{E})\ dV, \:\:\mbox{where}\:\: 
D(\vec{E})= \vec{E}:\vec{S}(\vec{E})\end{equation}
Substitution of the resulting expression for $\mathcal{D}[\lambda\vec{v}]$ of into the integrand in (\ref{funspace}) gives
\begin{equation} \Psi[\vec{v}]= \int_\Omega \psi(\vec{E})\ dV, \:\:\mbox{since}\:\:\psi(\vec{E})=\int_0^1 D(\lambda\vec{E})\lambda^{-1}d\lambda, \end{equation}
By taking the domain $\mathcal{R}= \Omega_F\cup I$, one obtains by similar
reasoning the final relation in (\ref{diss}).  
It seems plausible that one might  be able further to establish a 
pseudo-linear form for the force appearing in (\ref{funspace}):\begin{equation} \vec{f}\{\vec{v}\}:=\vec{f}[\vec{v}](\vec{x}) =  \int_\mathcal{R} \boldsymbol{\mathcal{L}}[\vec{v}](\vec{x},\vec{y})\vec{v}(\vec{y})\ d\mathcal{R}(\vec{y}),\end{equation}  involving a positive-definite matrix $\boldsymbol{\mathcal{L}}[\vec{v}](\vec{x},
\vec{y})$. However, this is a matter requiring a more thorough mathematical treatment.

\bibliographystyle{apalike}
\bibliography{nneo24arxiv}

\end{document}